\documentstyle[aps,epsf]{revtex}
\tighten
\begin{document}
\draft

\title{Zero-Field Satellites of a Zero-Bias Anomaly} 
\author{V. M. Apalkov and M. E. Raikh}
\address{Department of Physics, University of Utah, Salt Lake City,
Utah 84112}

\maketitle
\begin{abstract}
Spin-orbit (SO) splitting, $\pm \omega_{\mbox{\tiny SO}}$,
of the electron Fermi surface
in two-dimensional systems manifests  itself
in the interaction-induced corrections to the tunneling density of
states, $\nu (\epsilon)$.
Namely, in the case of a smooth disorder, it gives rise to the
satellites  of a zero-bias anomaly at energies 
$\epsilon=\pm 2\omega_{\mbox{\tiny SO}}$.  Zeeman splitting,
$\pm \omega_{\mbox{\tiny Z}}$, in a weak parallel magnetic field
causes a narrow {\em plateau} of a width 
$\delta\epsilon=2\omega_{\mbox{\tiny Z}}$ 
at the top of each sharp satellite peak. 
%broadens the satellites. 
As $\omega_{\mbox{\tiny Z}}$ exceeds 
$\omega_{\mbox{\tiny SO}}$, the SO satellites cross over to the 
conventional narrow maxima at 
$\epsilon = \pm 2\omega_{\mbox{\tiny Z}}$ with SO-induced plateaus 
$\delta\epsilon=2\omega_{\mbox{\tiny SO}}$ at the tops.
\end{abstract}
%\vskip2pc]
%\maketitle
%\narrowtext

%$\Lambda${\Large $\Lambda$} {\Large $\Pi $}

\noindent{\em Introduction.}\ A zero-magnetic-field 
splitting\cite{bychkov84} of the electron spectrum in 
two-dimensional systems has its origin in the spin-orbit (SO) 
coupling. The effect
attracts a steady interest especially since the proposal\cite{datta90}
 to utilize it in a
spin-transistor device.  This
proposal was substantiated by the recent experimental
demonstration\cite{nitta97,engels97} that the magnitude of the
splitting, $2\omega_{\mbox{\tiny SO}}$, can be controlled by 
the gate voltage.

A direct consequence of a finite zero-field splitting is the beating
pattern in the Shubnikov-de Haas oscillations\cite{luo90}.  These
beats remain the prime tool for the experimental determination of
$\omega_{\mbox{\tiny SO}}$.  However, the reliable results were
obtained only for electrons in narrow-gap heterostructures, where 
the splitting is relatively strong (several {\em meV}\cite{sato01}).
In the structures with small $\omega_{\mbox{\tiny SO}}$ (e.g.,
$\omega_{\mbox{\tiny SO}} \sim 0.5$ {\em meV} in {\em GaAs}--based
heterostructures\cite{ramvall97}), inferring the splitting from
beating pattern becomes ambiguous, since it requires extrapolation 
of the beats period to a zero magnetic field.

In addition to the beats of the Shubnikov-de Haas oscillations,
a number of more delicate effects
that are due {\em exclusively} to the finite zero-field splitting of
the conduction band have been observed experimentally
\cite{dresselhaus92,jusserand92,knap96,morpurgo98,folk01}.
Still, extracting small $\omega_{\mbox{\tiny SO}}$
values from the data remains a problem.

In the present paper we identify another prominent manifestation 
of the zero-field splitting. Namely, we demonstrate that SO coupling
causes anomalies in the tunneling conductance,
$g(V)$, at a {\em finite} bias, $V$.  The positions of anomalies
$V=\pm 2\omega_{\mbox{\tiny SO}}$ 
reveal directly the splitting magnitude. They  emerge as satellites 
of the conventional zero-bias anomaly at $V=0$. The latter
is the result of electron-electron interactions modified by the
disorder\cite{altshuler79}.
To the best of our knowledge, these SO-induced satellites
represent the first case when a 
{\em disorder--related} effect in transport allows to infer the 
parameter of the {\em intrinsic} electronic spectrum.
We note that, in the experiment, very fine features 
(on the scale of
$\epsilon \sim 0.1$ {\em meV}\cite{wu96,butko99,chan97})
in the 
tunneling density of states, $\nu(\epsilon)$, can be resolved
upon the analysis of $g(V)$ curves at low 
temperatures $T\ll V$.

In two dimensions, the interaction-induced
correction, $\delta \nu (\epsilon)$, to the free-electron density of
states, $\nu_0$, behaves as \cite{altshuler80}
$\delta \nu (\epsilon) \propto \ln(\vert \epsilon\vert \tau)$,
for $\epsilon \ll \tau^{-1}$ (diffusive regime), where $\tau$
is the scattering time. This simple logarithmic form applies in the
presence of a tunneling electrode which causes  a long-distance 
cut-off of the Coulomb interactions\cite{zyuzin81}. 
Zeeman splitting of the electron spectrum in a parallel magnetic 
field leads to  additional
anomalies 
$\delta\nu (\epsilon)\propto \ln(\vert\epsilon\pm 
2\omega_{\mbox{\tiny Z}} \vert \tau)$
\cite{altshuler85}.
Remarkably, these anomalies emerge even in the weak-field limit, 
when
$\omega_{\mbox{\tiny Z}} \ll \tau^{-1}$. 
This is because the scattering by the 
impurity potential does not mix the spin-split subbands. 
Thus, the wave functions
of two particles with opposite spins that differ in energy
by $2\omega_{\mbox{\tiny Z}} $ are strictly identical. 
Naturally, the SO {\em scattering} from the impurities suppresses 
the anomalies at $\epsilon =\pm 2\omega_{\mbox{\tiny Z}}$. 

The situation with SO coupling is somewhat opposite. In this case,
the disorder potential causes a mixing of the SO subbands.
If the disorder is short-ranged, each scattering act
results in the momentum transfer of the order of the Fermi wave 
vector, $k_{\mbox{\tiny F}}$. Thus, for weak SO coupling,
$\omega_{\mbox{\tiny SO}}\tau \ll 1$, the mixing is strong, and
the SO satellites of the zero-bias anomaly are smeared out.
If, however,
the disorder is smooth,
the  momentum transfer is much smaller than $k_{\mbox{\tiny F}}$.
In this generic regime of a small-angle scattering, the SO subbands
are almost decoupled\cite{dmitriev01} (see Fig.~1). 
As a result, the same reasoning that leads to the Zeeman 
satellites  applies, so that  
 the SO satellites  are pronounced even if
$\omega_{\mbox{\tiny SO}}$ is much smaller than $\tau^{-1}$.
Analogously, with a smooth disorder, in the presence of
a perpendicular magnetic field, electronic states separated in
energy by a cyclotron quantum turn out to be strongly correlated
\cite{rudin98}. Consequently, the interaction-induced 
$\delta\nu(\epsilon)$ is an oscillatory function 
of energy\cite{rudin97'} at low (compared to $\tau^{-1}$)
 cyclotron frequencies. 
Under the condition of a small-angle scattering, the dominant
contribution to $\delta\nu$ comes from the Hartree correction
\cite{rudin97'}. 

Below we also study the evolution
of the SO satellites with a parallel magnetic field. 
Due to a a non-trivial interplay between SO coupling and 
Zeeman splitting \cite{chen99',chen00,aleiner01},
this evolution
is rather peculiar. Namely, weak magnetic field causes  
{\em a flat top} of the SO satellite within a narrow interval
$\delta\epsilon=2\omega_{\mbox{\tiny Z}}\ll 
\omega_{\mbox{\tiny SO}} $. 
Conversely, in a strong magnetic field, the Zeeman satellite 
\cite{altshuler85} acquires a flat top of a width
$\delta\epsilon=2\omega_{\mbox{\tiny SO}}\ll 
\omega_{\mbox{\tiny Z}} $.

\noindent{\em Basic Equations.}\
Microscopic origin of the SO coupling
for two-dimensional electrons can be either
inversion asymmetry of the host crystal\cite{dresselhaus55} or
the  confinement potential asymmetry\cite{bychkov84}. Unless two
effects are comparable in strength\cite{pikus95}, we can choose the
form $\alpha (\bbox{k}
  \times \hat{\bbox{\sigma }}) \bbox{n}$
  for the SO Hamiltonian\cite{bychkov84},
where $\alpha$ is the coupling constant, $\hat{\bbox{\sigma}}$ 
is the spin operator, and $\bbox{n}$ is the unit vector 
normal to the two-dimensional plane.
In a parallel magnetic field, that induces the Zeeman splitting 
$2\omega_{\mbox{\tiny Z}}$, the Hamiltonian of a free electron 
has a form
\begin{equation}
\label{simplest}
\hat{H} = \frac{\hbar ^2 k^2}{2m} + \alpha (\bbox{k}
  \times \hat{\bbox{\sigma }}) \bbox{n} + \omega_{\mbox{\tiny Z}} \hat{\sigma} _x,
\end{equation}
where $m$ is the electron mass.
The spectrum of the Hamiltonian Eq.~(\ref{simplest}) 
represents two branches, so that in the vicinity of the Fermi 
surface we have 
\begin{equation}
E_{\mu } (k) = E_{F} +\epsilon _{\mu }(\bbox{k}) ,
\end{equation}
 where
\begin{equation}
\epsilon _{\mu } (\bbox{k}) = \hbar v_{F} (k - k_{F})  +
\mu \Delta (\bbox{k}) ,
\label{ep_1}
\end{equation}
and
\begin{equation}
\Delta (\bbox{k}) = \sqrt{ \omega _{\mbox{\tiny SO}} ^2 +\omega_{\mbox{\tiny Z}}^2 +
 2  \omega _{\mbox{\tiny SO}}  \omega_{\mbox{\tiny Z}} \sin \phi _{\bbox{k}}} .
\label{ep_2}
\end{equation}
Here $v_{\mbox{\tiny F}} = \hbar k_{\mbox{\tiny F}}/m$ is the Fermi velocity, 
$\mu =\pm 1$ is the branch index, 
$ \omega _{\mbox{\tiny SO}}=2\alpha k_{\mbox{\tiny F}}$ 
is the SO splitting, and
 $\phi _{\bbox{k}}$ is the azimuthal angle of $\bbox{k}$.
It is convenient to rewrite the diagonalized Hamiltonian 
Eq.~(\ref{simplest}) in the form
\begin{equation}
\label{hamil}
\hat{H}=\sum _{\mu }
E_{\mu } (\bbox{k}) \mbox{\Large $\hat{\Lambda }$} _{\mu } (\bbox{k}),
\end{equation}
where the projection operators $\mbox{\Large $\hat{\Lambda }$} _{\mu }$
are defined as\cite{chen99'}
\begin{equation}
\mbox{\Large $\hat{\Lambda }$} _{\mu } ( \bbox{k} ) =
 \frac{1}{2} \left( \begin{array}{cc}
   1 &   \mu \exp (-i \varphi _{\bbox{k}} )  \\
   - \mu \exp (i \varphi _{\bbox{k}} ) & 1
 \end{array} \right).
 \label{lambda1}
\end{equation}
The angle $\varphi _{\bbox{k}}$ is related to the 
azimuthal angle $\phi _{\bbox{k}}$ as follows
\begin{equation}
\tan \varphi _{\bbox{k}} =\tan \phi _{\bbox{k}} +
\frac{\omega_{\mbox{\tiny Z}}}{ \omega _{\mbox{\tiny SO}}\cos \phi _{\bbox{k}}} .
\label{angle}
\end{equation}
In the presence of the disorder, the electron scattering time is 
determined by  two processes, namely,  intra-subband scattering 
\begin{equation}
\frac{1}{\tau _{\mu \mu }(\bbox{k})} = \int \frac{d \bbox{p}}{(2 \pi )^2}
 \mbox{Tr} \left( \mbox{\Large $\hat{\Lambda }$} _{\mu } ( \bbox{k} )
                   \mbox{\Large $\hat{\Lambda }$} _{\mu } ( \bbox{p} )  \right)
\mbox{\Large $S$} \mbox{\Large $($} |\bbox{k} - \bbox{p}| \mbox{\Large $)$} ~ 
\delta \!\left( E_{\mu } (\bbox{k}) -
                  E_{\mu }(\bbox{p}) \right),
\label{tau_1}
\end{equation}
and inter-subband scattering
\begin{equation}
\frac{1}{\tau _{\mu, -\mu }(\bbox{k})} = \int \frac{d \bbox{p}}{(2 \pi )^2}
 \mbox{Tr} \left( \mbox{\Large $\hat{\Lambda }$} _{\mu } ( \bbox{k} )
                   \mbox{\Large $\hat{\Lambda }$} _{-\mu } ( \bbox{p} )  \right)
 \mbox{\Large $S$}\mbox{\Large $($} 
|\bbox{k} - \bbox{p}|  \mbox{\Large $)$} ~\delta \! \left( E_{\mu } (\bbox{k}) -
                  E_{-\mu }(\bbox{p}) \right) =
                   \frac{1}{\tau _{int}(\bbox{k})},
\label{tau_2}
\end{equation}
where $ \mbox{\Large $S$} \mbox{\Large $($} k \mbox{\Large $)$} $ is 
the Fourier transform of the correlator of the random potential. 
Our assumption that the disorder is smooth can be quantitatively 
expressed as $\lambda _{\bbox{k},\bbox{p}} \ll 1$, where 
the parameter $\lambda _{\bbox{k},\bbox{p}}$ is defined as
\begin{equation}
\lambda _{\bbox{k},\bbox{p}}= \mbox{Tr} 
  \left( \mbox{\Large $\hat{\Lambda }$} _{\mu } ( \bbox{k} )
                   \mbox{\Large $\hat{\Lambda }$} _{-\mu } ( \bbox{p} )  \right)
              = \frac{1-\cos (\varphi _{\bbox{k}} - \varphi _{\bbox{p}})}{2}
          \approx \frac{ ( \varphi _{\bbox{k}} - \varphi _{\bbox{p}})^2}{4}.
\label{Tr_1}
\end{equation}
Correspondingly, $\mbox{Tr}
                \left( \mbox{\Large $\hat{\Lambda }$} _{\mu } ( \bbox{k} )
             \mbox{\Large $\hat{\Lambda }$} _{\mu } ( \bbox{p} )  \right) =
          1 - \lambda _{\bbox{k},\bbox{p}}$ is close to unity.
Using Eq. (\ref{angle}), the parameter
 $\lambda _{\bbox{k}\bbox{p}}$ can be expressed through  
the angles $\phi _{\bbox{k}}$ and $\phi _{\bbox{p}}$ as
\begin{equation}
 \lambda _{\bbox{k}\bbox{p}} \approx \frac{\omega _{\mbox{\tiny SO}}^2
  ( \omega _{\mbox{\tiny SO}} +\omega _{\mbox{\tiny Z}} \sin \phi _{\bbox{k}} )^2 }
  {4 \Delta ^4(\bbox{k}) }
    \mbox{\Large $($} \phi _{\bbox{k}} - \phi _{\bbox{p}}  \mbox{\Large $)$}^2 .
\label{lambda_1}
\end{equation}
From Eqs.~(\ref{tau_1}), (\ref{tau_2}) we get the final expression for
the scattering time 
\begin{equation}
\frac{1}{\tau } = \frac{1}{\tau _{\mu \mu}} +  \frac{1}{\tau _{\mu, -\mu}}
 = \frac{m}{2\pi } \int d\varphi _{\bbox{p}} ~
\mbox{\Large $S$} \mbox{\Large $($}k_{\mbox{\tiny F}} 
   \varphi _{\bbox{p}}\mbox{\Large $)$},
\end{equation}
where we assumed that $\omega _{\mbox{\tiny SO}}$,
$\omega_{\mbox{\tiny Z}} \ll E_F$.

As it was discussed above, the satellite anomaly in 
$\delta \nu (\epsilon )$ originates from the 
Hartree correction. In the case of two subbands, 
the  expression for the energy-dependent part 
of the Hartree correction has the form
\begin{eqnarray}
\frac{\delta \nu (\epsilon )}{\nu _0}  & = & \frac{1}{2\pi }
\frac{\partial }{\partial \epsilon } \mbox{Re}
\int_{\epsilon } ^{\infty }\!\!\!\! d\omega \int \frac{d \bbox{q}}{(2\pi )^2}
\int  \frac{d \bbox{p}}{(2\pi )^2}
\int  \frac{d \bbox{p}^{\prime }}{(2\pi )^2}
   V(\bbox{p} - \bbox{p}^{\prime} ) 
\mbox{\Large $\Gamma$} ^{++}_{--}(\bbox{p}, \bbox{p}^{\prime },
   \bbox{q}, \omega ) \left(1- \lambda _{\bbox{p}+\bbox{q},
      \bbox{p}^{\prime }+\bbox{q}}  \right)
       \left(1- \lambda _{\bbox{p},
      \bbox{p}^{\prime }}  \right)    \nonumber \\
& & \times  G_{1}^{R} (\epsilon +\omega, \bbox{p}+\bbox{q})
 G_{1}^{R} (\epsilon +\omega, \bbox{p}^{\prime }+\bbox{q})
 G_{-1}^{A} (\epsilon , \bbox{p})
 G_{-1}^{A} (\epsilon , \bbox{p}^{\prime }),
 \label{H_1}
\end{eqnarray}
where $V(\bbox{p})$ is the Fourier transform of the 
screened electron-electron interaction. The retarded and 
advanced Green functions are the following
matrices
\begin{equation}
\hat{G}^{R,A} (\epsilon , \bbox{p}) = \sum _{\mu }
   \frac{ \mbox{\Large $\hat{\Lambda }$} _{\mu } (\bbox{p}) }
        { \epsilon - \epsilon _{\mu } (\bbox{p}) \pm  \frac{i}{2\tau } }
        = \sum _{\mu }
    \mbox{\Large $\hat{\Lambda }$} _{\mu } (\bbox{p})
     G^{R,A} _{\mu } (\epsilon , \bbox{p}) .
\label{green}
\end{equation}
The two-particle vertex function, $\mbox{\Large $\Gamma $}^{++}_{--} $, that 
is responsible for satellites,
is determined from the standard Dyson-type equation with a kernel
\begin{eqnarray}
\mbox{\Large $K$}(\bbox{p}, \bbox{p}_1,\bbox{q},\omega ) & = &
 e^{i (\varphi _{\bbox{p}}-\varphi _{\bbox{p}_1})}
  \left( 1 -  \frac{1}{2} \lambda _{\bbox{p}+\bbox{q},\bbox{p}_1+\bbox{q}}
           -  \frac{1}{2} \lambda _{\bbox{p},\bbox{p}_1} \right)
   e^{-i (\varphi _{\bbox{p}+\bbox{q}}-\varphi _{\bbox{p}_1+\bbox{q}})} 
                                                          \nonumber \\
 & & \times   \left\langle  G_{1}^{R} (\epsilon +\omega, \bbox{p}_1+\bbox{q})
 G_{-1}^{A} (\epsilon , \bbox{p}_1) \right\rangle 
\mbox{\Large $S$} \mbox{\Large $($} k_{\mbox{\tiny F}} |\phi _{\bbox{p}} -
  \phi _{\bbox{p}_1}| \mbox{\Large $)$} ,
\label{kernel}
\end{eqnarray}
where the first three factors originate from the overlap integrals. 
With a bare spectrum Eqs.~(\ref{ep_1}), (\ref{ep_2}) 
the product $ G_{1}^{R} (\epsilon +\omega, \bbox{p}_1+\bbox{q})
 G_{-1}^{A} (\epsilon , \bbox{p}_1) $ averaged over the disorder
 is given by
\begin{equation}
 \left\langle  G_{1}^{R} (\epsilon +\omega, \bbox{p}_1+\bbox{q})
 G_{-1}^{A} (\epsilon , \bbox{p}_1) \right\rangle =
    \frac{m \tau }{2 \pi } \left[ 1 - i \mbox{\Large $($} \omega -\Delta (\bbox{p})
       -\Delta (\bbox{p}+\bbox{q}) \mbox{\Large $)$} \tau +
         i \hbar q v_{\mbox{\tiny F}}   \tau \cos(\phi _{\bbox{p}} - 
      \phi _{\bbox{q}}) \right]^{-1}.
\label{av}
\end{equation}
Small-angle scattering implies that a 
typical $q\ll k_{\mbox{\tiny F}}$. This allows to 
set in Eq.~(\ref{kernel}) $\Delta ( \bbox{p}+\bbox{q} ) \approx \Delta (\bbox{p} ) $,
$\lambda _{\bbox{p}+\bbox{q},\bbox{p}_1+\bbox{q}} \approx \lambda
_{\bbox{p},\bbox{p}_1}$, and $\phi _{\bbox{p} +\bbox{q}} \approx \phi _{\bbox{p}}$. 
Then the kernel Eq.~(\ref{kernel}) simplifies to 
\begin{equation}
\mbox{\Large $K$} (\bbox{p}, \bbox{p}_1,\bbox{q},\omega ) = 
                                 \left(  \frac{m \tau }{2 \pi } \right)
  \frac{ (1-\lambda _{\bbox{p},\bbox{p}_1})\mbox{\Large $S$} 
 \mbox{\Large $($} k_{\mbox{\tiny F}}|\phi _{\bbox{p}} - 
      \phi _{\bbox{p}_1}|\mbox{\Large $)$}  }
   {1 - i \mbox{\Large $($} \omega -2\Delta (\bbox{p}) \mbox{\Large $)$} \tau +
    i \hbar q v_{\mbox{\tiny F}} \tau \cos(\phi _{\bbox{p}} - \phi _{\bbox{q}}) }.
 \label{kernel_1}
\end{equation}
At this point we note that, upon integration over $\phi _{\bbox{p}_1}$, 
the product $(1-\lambda _{\bbox{p},\bbox{p}_1}) \mbox{\Large $S$}
\mbox{\Large $($} k_{\mbox{\tiny F}}|\phi _{\bbox{p}} - \phi _{\bbox{p}_1}| 
      \mbox{\Large $)$} $ is
 proportional to $\left(\tau ^{-1} - \tau ^{-1} _{int}(\bbox{p}) \right)$,
where $\tau _{int}$ is the inter-subband scattering time (\ref{tau_2}). Then,
 in the diffusive regime, 
 $ \mbox{\Large $($} \omega -2\Delta (\bbox{p}) \mbox{\Large $)$}\tau\ll 1 $,
 the solution of the Dyson equation for the vertex function 
 $\mbox{\Large $\Gamma $} ^{++}_{--}(\bbox{p}, \bbox{p}^{\prime }, 
\bbox{q}, \omega) $
reads
\begin{equation}
\mbox{\Large $\Gamma$} ^{++}_{--}(\bbox{p}, \bbox{p}^{\prime }, \bbox{q}, \omega) =
 \frac{ \mbox{\Large $S$}\mbox{\Large $($} 
                 k_{\mbox{\tiny F}}|\phi _{\bbox{p}} - \phi _{\bbox{p} ^{\prime }}|
  \mbox{\Large $)$}}
 { -i \mbox{\Large $($} \omega - 2 \Delta (\bbox{p})
 \mbox{\Large $)$} \tau + D q^2 \tau 
     + \tau / \tau _{int}(\bbox{p}) },
\label{Gamma_0}
\end{equation}
where we have neglected a weak anisotropy of the diffusion coefficient
$D =  v_{\mbox{\tiny F}}^2 \tau_{tr} /2$.

In principle, the Dyson equation for $\mbox{\Large $\Gamma$} ^{++}_{--}$ couples
(weakly) this function to the other vertex functions, {\em i.e.} 
$\mbox{\Large $\Gamma$} ^{--}_{++}$. This coupling would be 
important in the domain $\omega _{\mbox{\tiny SO}} \ll
\tau _{int}^{-1}$. In our case, $\omega _{\mbox{\tiny SO}}
\tau _{int} \gg 1$, this coupling can be neglected.

Substituting Eq.~(\ref{Gamma_0}) into Eq.~(\ref{H_1}) and performing
integration over $p$ and $p^{\prime }$, we obtain
\begin{equation}
\frac{\delta \nu (\epsilon )}{\nu _0}  = \frac{\tau \nu _0}{4\pi }~
\mbox{Re} \int  _0 ^{1/v_{\mbox{\tiny F}}\tau} dq~q
\int _0^{2 \pi} \frac{d \phi_{\bbox{p}}}{2\pi }
\int _0^{2 \pi} \frac{d \phi_{\bbox{p}^{\prime }}}{2\pi }
 \frac{ V(k_{\mbox{\tiny F}}|\phi _{\bbox{p}} - \phi _{\bbox{p} ^{\prime }}|)
        \mbox{\Large $S$}\mbox{\Large $($} k_{\mbox{\tiny F}}|\phi _{\bbox{p}} - 
                                    \phi _{\bbox{p} ^{\prime }}|\mbox{\Large $)$}}
 {-i \mbox{\Large $($}\epsilon - 2 \Delta (\bbox{p}) \mbox{\Large $)$} + D q^2
     + \tau ^{-1}_{int}(\bbox{p}) }.
 \label{H_2}
\end{equation}
The fact that characteristic  $ \phi _{\bbox{p}} - \phi _{\bbox{p} ^{\prime }}$
is small allows to set 
$  V (k_{\mbox{\tiny F}}|\phi _{\bbox{p}} - 
\phi _{\bbox{p} ^{\prime }}|)=V(0)$, where 
 $V(0) = 1/\nu _0$ (static screening). Then integration over  
$\phi _{\bbox{p} ^{\prime }}$ yields $1/(m\tau )$. Finally we obtain
\begin{equation}
\frac{\delta \nu (\epsilon )}{\nu _0} 
 = - \left(\frac{1}{16 \pi E_F \tau _{tr} }\right)
  {\cal L}(\epsilon ),
 \label{H_3}
\end{equation}
where the energy-dependent factor, ${\cal L} (\epsilon )$, is defined as 
\begin{equation}
{\cal L} (\epsilon )= \int _0^{2 \pi} \frac{d \phi }{2\pi }
\ln \left[ \mbox{\Large $($} \epsilon - 2\Delta (\phi )\mbox{\Large $)$}^2  \tau^2 +
 \frac{\tau ^2}{\tau _{int}^2 (\phi) }
  \right].
\label{L}
\end{equation}
We see that the energy cut-off in Eq.~(\ref{L}) is determined by $\tau _{int}^{-1}
\ll \tau ^{-1}$. 
Note, that inter-subband scattering time $\tau _{int}$ can be conveniently 
expressed through the conventional transport relaxation time $\tau _{tr}$.  
Using Eqs.~(\ref{tau_2}), (\ref{lambda_1}) we obtain 
\begin{equation}
\tau _{int} (\phi ) = 
  2 \tau_{tr} \frac{\left(\omega_{\mbox{\tiny Z}}^2 +
\omega_{\mbox{\tiny SO}}^2 + 2 \omega_{\mbox{\tiny Z}}\omega_{\mbox{\tiny SO}}
 \sin \phi \right)^2}{\omega_{\mbox{\tiny SO}}^2 \left(
 \omega_{\mbox{\tiny SO}}+\omega_{\mbox{\tiny Z}}\sin \phi \right)^2}.
\label{tau5}
\end{equation}
Equations (\ref{H_3})--(\ref{tau5}) constitute our main result.
In principle, the exchange correction to 
$\nu (\epsilon )$ also yields the anomaly at 
$\epsilon \approx 2 \Delta $. However, the exchange term is
suppressed since it contains an extra factor $\lambda  \sim 
\tau/\tau _{tr}$ which comes from overlap integrals between 
different branches. 
Note also, that Eqs.~(\ref{H_3})--(\ref{tau5}) can be easily 
modified  to the case
when the crystalline anisotropy term \cite{dresselhaus55},
$\beta (\hat{\sigma }_x k_x -\hat{\sigma }_y k_y)$,  is
present in the  Hamiltonian (\ref{simplest}) alongside with SO-term 
\cite{bychkov84}. Modification reduces to the replacement 
$\Delta _{\mbox{\tiny Z}}$ by $\beta k_{\mbox{\tiny F}}$,
and $\phi $ by $2\phi $ \cite{chen00}.

\noindent {\em Analysis of the anomaly.}\ 

({\em i}) Zero-field limit: $\omega_{\mbox{\tiny Z}} \ll 1/ \tau _{tr} \ll
  \omega _{\mbox{\tiny SO}} $. \\ The shape of the satellite peak in
$\delta \nu (\epsilon )$ is given by
\begin{equation}
\frac{\delta \nu (\epsilon )}{\nu _0} = -\left( \frac{1}{16 \pi E_F \tau _{tr}}
  \right) \ln \left( (\epsilon - 2 \omega _{\mbox{\tiny SO}} )^2 \tau^2 +
    \frac{\tau ^2}{4\tau _{tr}^2}  \right).
\label{i}
\end{equation}
The peak is well pronounced when $\omega _{\mbox{\tiny SO}} \tau _{tr} \gg 1$.

({\em ii}) Intermediate fields: 
 $1/\tau _{tr} \ll \omega_{\mbox{\tiny Z}} \ll  \omega _{\mbox{\tiny SO}}$. \\
The broadening of the SO-satellite peak is determined by the angular
dependence of $\Delta (\phi )$. The integral ${\cal L}(\epsilon )$
can be evaluated in this limit, yielding
\begin{equation}
{\cal L} (\epsilon ) = 2\ln \! \mbox{\huge $|$}
 \!\!\left( \frac{\epsilon }{2} -  \omega _{\mbox{\tiny SO}}\right) \tau
 + \sqrt{
 \left( \frac{\epsilon}{2} -
   \omega _{\mbox{\tiny SO}} \right)^2\tau^2 
 - \omega_{\mbox{\tiny Z}}^2 \tau^2 
  } 
\mbox{\huge $|$}  .
 \label{ii}
\end{equation}
Remarkably, within a domain 
$|\epsilon - 2 \omega _{\mbox{\tiny SO}} |
\leq  \omega_{\mbox{\tiny Z}}$, there is a {\em plateau} in
 $\delta \nu (\epsilon )$,
{\em i.e.} within this domain ${\cal L} (\epsilon ) \equiv 
\ln \omega_{\mbox{\tiny Z}} \tau $.  

({\em iii}) Strong fields: $1/\tau _{tr} \ll  \omega _{\mbox{\tiny SO}} \ll
\omega_{\mbox{\tiny Z}} $. \\
  In this case, conversely, SO coupling determines the
shape of the Zeeman satellite
\begin{equation}
{\cal L} (\epsilon ) = 2\ln \! \mbox{\huge $|$}
 \!\!\left( \frac{\epsilon }{2} -  \omega _{\mbox{\tiny Z}}\right) \tau
 + \sqrt{
 \left( \frac{\epsilon}{2} -
   \omega _{\mbox{\tiny Z}} \right)^2\tau^2 
 - \omega_{\mbox{\tiny SO}}^2 \tau^2 
  } 
\mbox{\huge $|$}   .
 \label{iii}
\end{equation}
Again a plateau at $|\epsilon - 2 \omega_{\mbox{\tiny Z}}| \leq 
 \omega _{\mbox{\tiny SO}} $ emerges at the top of the 
 Zeeman satellite.

Typical examples of the energy dependence of $\delta\nu(\epsilon )$,
obtained by numerical integration of Eq.~(\ref{L}),  are shown in 
Fig.~2. They illustrate the successive broadening and then narrowing
of the satellite peak with increasing magnetic field. 

\noindent {\em Conclusion.}\ The above consideration was
restricted to the diffusive regime $\epsilon \sim
 \omega _{\mbox{\tiny SO}}  \ll 1/\tau$. It is 
known however\cite{rudin97,khveshchenko98,mishchenko01}, that the 
conventional diffusive zero-bias
anomaly 
persists at 
high electron energies $\epsilon \gg 1/\tau$ (ballistic regime).
The question
 arises whether the SO anomaly survives in the
 ballistic regime. 
We will discuss this case qualitatively.
 Without SO coupling, the physical mechanism responsible
 for the
 formation of the zero-bias anomaly is combined scattering of
a probe  electron  from an isolated impurity and a perturbation
 of the electron density caused by this impurity. 
The latter
perturbation falls off with distance as 
$\sin (2k_{\mbox{\tiny F}} r)/r^2$
(Friedel oscillation). Then the anomaly emerges as a result of  the 
Bragg
backscattering from this almost periodic potential profile.
In the presence of SO coupling, a single impurity induces
Friedel oscillation with {\em three} wave vectors\cite{chen99}, 
namely
$2k_{\mbox{\tiny F}}$, $2k_{\mbox{\tiny F}} \pm 2 
\omega _{\mbox{\tiny SO}} /( v_{\mbox{\tiny F}})$.
Then the wave vector for an electron with chirality $+1$ and
energy $2\omega _{\mbox{\tiny SO}} $ above the Fermi
level will satisfy the Bragg condition for the Friedel
oscillation created by electrons with chirality $-1$. This is
illustrated in Fig.~1. The efficiency of the Bragg scattering, and,
hence, the anomaly at $\epsilon = 2  \omega _{\mbox{\tiny SO}}$,
is suppressed in the absence of magnetic field due to the fact that,
for a given chirality,
the spinors corresponding to the wave vectors $\bbox{k} $ and
$-\bbox{k}$ are {\em orthogonal to each other}.
A transparent underlying physics of the ballistic zero-bias anomaly
suggests that electrons with both chiralities can experience 
Bragg scattering from the SO-specific $2k_{\mbox{\tiny F}}$--oscillation
\cite{chen99}. 
These resonances give rise to yet additional weak anomalies 
at energies $\epsilon=\pm \omega _{\mbox{\tiny SO}}$,  
that are absent in the diffusive regime.

Note in conclusion, that experimental studies \cite{coleridge91}
indicate that even in moderate quality {\em GaAs}--based samples with
mobility $\sim 2\cdot 10^5$ cm$^2$/V$\cdot $s (as in \cite{chan97})
the typical value of the ratio $\tau /\tau_{tr} $ is $0.1$.
Then for $\omega _{\mbox{\tiny SO}} =0.5 meV$ the parameter
$\omega _{\mbox{\tiny SO}} \tau_{tr} \approx 6.6$.

As a final remark, SO coupling for two-dimensional holes 
is much stronger that for electrons. Therefore, the
satellites of the zero-bias anomaly can be expected in the hole
samples too. However, due to the warp of the valence band spectrum
caused by the crystalline
anisotropy,  the subband splitting depends on the direction of the
hole wave vector. This would lead to the smearing of the satellites.

\noindent {\em Acknowledgments}.\ The authors are grateful to
R. Ashoori for a valuable discussion.

\begin{figure}
\centerline{
\epsfxsize=4.2in
\epsfbox{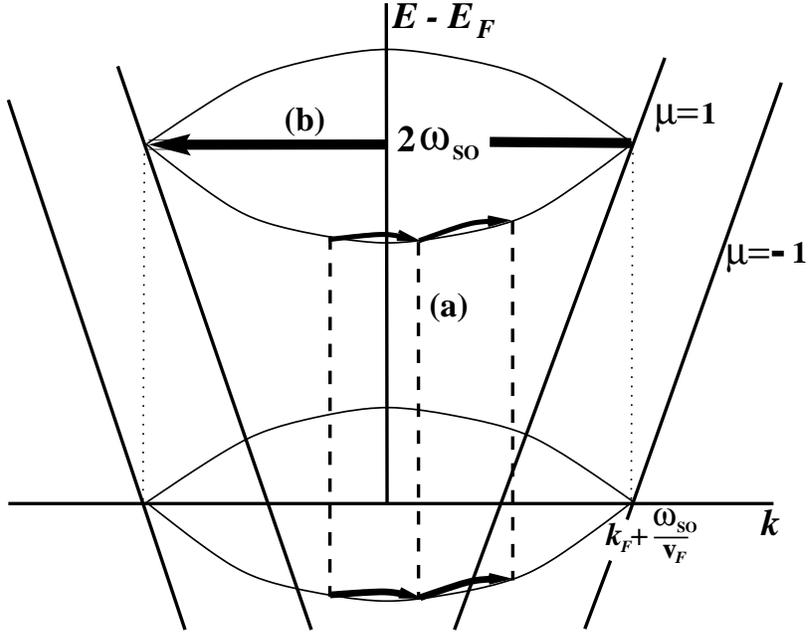}}
\vspace*{0.1in}
\protect\caption[sample]
{\sloppy{ Schematic illustration of the processes responsible 
for satellite anomaly at $\epsilon = 2 \omega _{\mbox{\tiny SO}}$
in the diffusive (a) and ballistic (b) regimes. 
}}
\label{figone}
\end{figure}

\begin{figure}
\centerline{
\epsfxsize=4.2in
\epsfbox{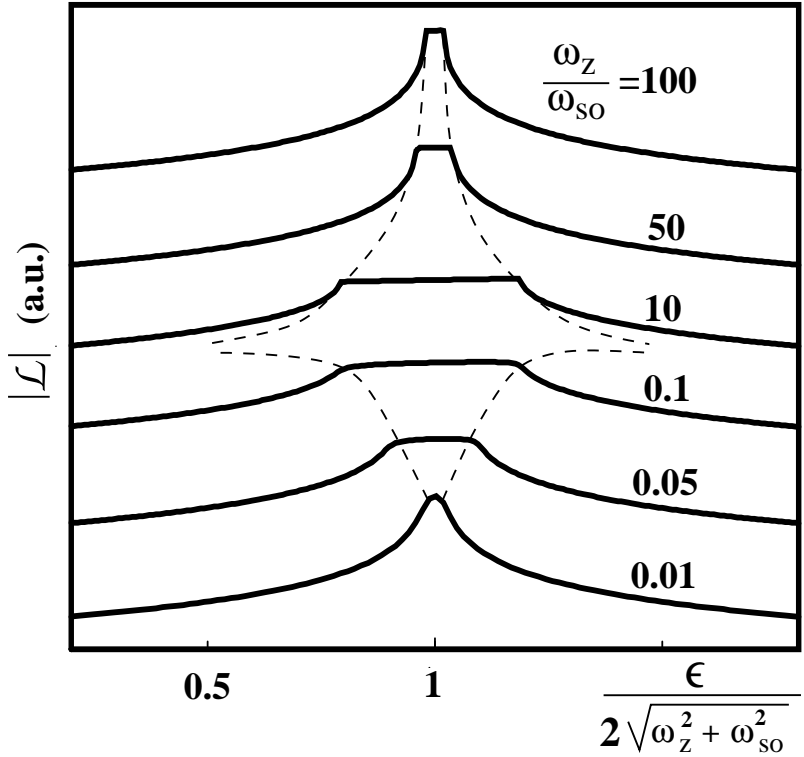}}
\vspace*{0.1in}
\protect\caption[sample]
{\sloppy{ Normalized correction, $\delta \nu (\epsilon )$, 
calculated from Eqs.~(\ref{L}), (\ref{tau5}) for 
$\omega _{\mbox{\tiny SO}}\tau _{tr}=10$
is plotted versus dimensionless energy $\epsilon /
2(\omega _{\mbox{\tiny SO}}^2+\omega _{\mbox{\tiny Z}}^2)^{1/2}$
for various ratios 
$\omega _{\mbox{\tiny Z}}/\omega _{\mbox{\tiny SO}}$. For
convenience different curves are shifted along the vertical 
axis.  
}}
\label{figtwo}
\end{figure}

\end{document}